
\def\mline{\hbox to 5in{\vrule width5in height0.6pt depth0.0pt}}
\def \hm1{h^{-1}}
\def \ie {{\it i.e. }}
\def \eg {{\it e.g. }}
\def \etal {{\it et al.\/}}
\def \nbar {{\overline n}}
\def \br {{\bf r}}
\def \bk {{\bf k}}
\def\lsim{\mathrel{\mathchoice {\vcenter{\offinterlineskip\halign{\hfil
$\displaystyle##$\hfil\cr<\cr\sim\cr}}}
{\vcenter{\offinterlineskip\halign{\hfil$\textstyle##$\hfil\cr<\cr\sim\cr}}}
{\vcenter{\offinterlineskip\halign{\hfil$\scriptstyle##$\hfil\cr<\cr\sim
\cr}}}
{\vcenter{\offinterlineskip\halign{\hfil$\scriptscriptstyle##$\hfil\cr<\cr
\sim\cr}}}}}
\def\gsim{\mathrel{\mathchoice {\vcenter{\offinterlineskip\halign{\hfil
$\displaystyle##$\hfil\cr>\cr\sim\cr}}}
{\vcenter{\offinterlineskip\halign{\hfil$\textstyle##$\hfil\cr>\cr\sim\cr}}}
{\vcenter{\offinterlineskip\halign{\hfil$\scriptstyle##$\hfil\cr>\cr\sim
\cr}}}
{\vcenter{\offinterlineskip\halign{\hfil$\scriptscriptstyle##$\hfil\cr>\cr
\sim\cr}}}}}

\magnification \magstep1
\baselineskip 14pt plus 2pt
\parindent=1cm

\vskip 3cm
\centerline{\bf POWER SPECTRUM ANALYSIS}
\centerline{\bf OF}
\centerline{\bf THREE-DIMENSIONAL REDSHIFT SURVEYS}
\vskip 0.5cm
\centerline{\bf Hume A. Feldman$^{1,a}$, Nick Kaiser$^{2,4,b}$ and John A.
Peacock$^{3,4,c}$}
\vskip 0.5cm
\centerline{1) Physics Dept., University of Michigan, Ann Arbor, MI 48109}
\centerline{2) CITA, University of Toronto, Ontario M5S 1A1, Canada}
\centerline{3) Royal Observatory, Blackford Hill, Edinburgh EH9 3HJ, UK}
\centerline{4) CIAR Cosmology Program}

\beginsection Abstract

We develop a general method for power spectrum analysis of three
dimensional redshift surveys. We present rigorous analytical estimates
for the statistical uncertainty in the power and we are able to derive
a rigorous optimal weighting scheme under the reasonable (and largely
empirically verified) assumption that the long wavelength Fourier
components are Gaussian distributed. We apply the formalism to the
updated 1-in-6 QDOT IRAS redshift survey, and compare our results to
data from other probes: APM angular correlations; the CfA and the
Berkeley 1.2Jy IRAS redshift surveys.  Our results bear out and
further quantify the impression from e.g.\ counts-in-cells analysis
that there is extra power on large scales as compared to the standard
CDM model with $\Omega h\simeq 0.5$.  We apply likelihood analysis
using the CDM spectrum with $\Omega h$ as a free parameter as a
phenomenological family of models; we find the best fitting parameters
in redshift space and transform the results to real space.  Finally,
we calculate the distribution of the estimated long wavelength power.
This agrees remarkably well with the exponential distribution expected
for Gaussian fluctuations, even out to powers of ten times the mean.
Our results thus reveal no trace of periodicity or other non-Gaussian
behavior.

\vskip 0.5cm
{\it Subject Heading:} cosmology: theory -- dark matter -- observation --
galaxies: clustering
-- distances and redshifts -- infrared: galaxies
\vskip 1cm
\centerline{Submitted to {\it The Astrophysical Journal}}

\mline
\noindent $^a$feldman@pablo.physics.lsa.umich.edu

\noindent $^b$kaiser@chipmunk.cita.utoronto.ca

\noindent $^c$jap@star.roe.ac.uk

\vfill\eject
\baselineskip 16pt plus 2pt

\beginsection 1. INTRODUCTION

The IRAS QDOT survey (see e.g. Efstathiou \etal\ 1990) was designed
for the study of large--scale structure, and has been analyzed
statistically in a number of ways: counts in cells (Efstathiou \etal\
1990) topology (Moore \etal\ 1992) etc.  It has also been used to
predict the peculiar motions of the local group (Rowan--Robinson
\etal\  1990; Strauss \etal\  1990; Strauss \etal\  1992) and of other
galaxies (Saunders \etal\ 1991) and thereby give an estimate of the
density of the universe.  Here we shall present the results of
power--spectrum analysis of this survey, preliminary results of which
have appeared elsewhere (Kaiser 1992; Feldman 1993).

The present study follows the program laid down by Peebles in his
pioneering series of papers (Yu \& Peebles 1969; Peebles 1973; Peebles
\& Hauser 1974) for statistical analysis of  galaxy catalogues via low
order correlation functions. The power spectrum $P(k)$ is the Fourier
transform of the spatial two--point function $\xi(r)$ and, as such,
can only provide a partial description of the nature of the
large--scale structure. While limited in information, the two--point
function does hold a rather special place: if the initial fluctuations
were Gaussian then the power spectrum provides a complete description
of the fluctuations.  It also provides a good starting point for
higher--order analyses and, in any case, our knowledge of the
two--point function is as yet far from complete, particularly on large
scales.

While Peebles emphasized the role of the power spectrum as a tool for
two--point analysis, most work on real data has tended to use the
auto--correlation function as the primary estimator.  Recently, the
power spectrum has made something of a comeback and has been applied
to the CfA survey(s) (Baumgart \& Fry 1991; Vogeley \etal\ 1992), to
pencil beam surveys (Broadhurst \etal\ 1990; see also Kaiser \&
Peacock 1991) and to both 1--in--6 0.6 Jy (Kaiser \etal\ 1991) and
complete 1.2--Jy surveys (Strauss \etal\ 1990; Fisher \etal\ 1993)
derived from Version 2 of the IRAS {\it point source catalogue}
(Chester, Beichmann \& Conrow 1987). Power spectrum analysis has also
been applied to radio galaxies (Webster 1977; Peacock \& Nicholson
1991), to clusters of galaxies (Peacock \& West 1992; Einasto, Gramann
\& Tago 1993) and to the Southern Sky Redshift Survey (Park, Gott \&
da Costa 1992).

Since $P(k)$ and $\xi(r)$ are Fourier transform pairs, one might
question what is gained by all this analysis. While complete knowledge
of the two--point function $\xi(r)$ is equivalent to complete
knowledge of the power spectrum $P(k)$ since they are a Fourier
transform pair, the same is not true of {\it estimates\/} $\hat
\xi$ and $\hat P$ derived from finite and noisy observational samples.
There may be benefits to be derived from having both estimates and,
depending on the details of the survey at hand, there will be relative
advantages of one method over the other.

Whereas the auto--correlation function measures the excess probability
of finding a pair of galaxies in two volumes separated by some
distance, the power spectrum directly measures the fractional density
contributions on different scales. This, in general, is a more natural
quantity, and one that is being supplied by theories that describe the
state of the Universe at early times \eg inflationary theories.
Furthermore the correlation function, especially on large scales, is
very sensitive to assumptions we make about the mean density $\bar n$
(the correlation function $\xi_{_{12}}+1\equiv n_{_1}n_{_2}/\bar n^2$
where $1,2$ refer to two volumes); in fact, the scales where $\xi\ll1$
are the ones we are most interested in when studying large--scale
clustering.  In contrast, the power spectrum, which gives us a direct
measurement of $\delta\rho/\rho$, scales as $\bar n^{-1}$ and so
determines the density field on large scales more robustly.
Uncertainty in our knowledge of the mean density may amplify the error
in the correlation function; however, the shape of the power spectrum
should be unaffected by incomplete knowledge of $\bar n$ which is
measured by the $k=0$ mode.

Another general advantage of the power spectrum derives from the fact
that the true $P(k)$ is a positive quantity. Thus, in interpreting an
estimate $\hat P(k)$ --- which will not necessarily be positive --- we
have important extra information at our disposal.  It is therefore
possible to recognize directly unphysical results which may indicate
some problem with data or analysis program. For the auto--correlation
function this information translates into an infinite number of
integral constraints: the integral of $\xi(\br)$ times
$\exp(i\bk\cdot\br)$ must be positive for every value of $\bk$, but
this seems relatively obscure.

A related advantage of the power--spectrum analysis involves the error
analysis. In general, the statistical uncertainty in any two--point
function involves the 4--point function which is at best poorly known.
An interesting error estimate --- and the relevant one if one is
interested in testing models like CDM --- is obtained if we assume
that the large--scale structure resembles a Gaussian random field.
When we take the Fourier transform of the galaxies in the survey we
are, roughly speaking, transforming the product of the infinite random
sea of density fluctuations $f(\br)$ with a window determined by the
survey geometry. The transform will therefore be the convolution of
the Fourier coefficients $f(\bk)$ with a `point--spread function'
(psf) which, for the IRAS survey, is a compact ball with width equal
to the inverse of the depth of the survey. For a Gaussian field, the
coefficients $f(\bk)$ are totally uncorrelated with one another, so,
aside from the a short range coherence imposed by the convolution,
estimates of the power will also be statistically independent. In the
correlation function representation the fluctuations in $\xi(\br)$ at
different $\br$ will be correlated in some rather complicated way.

These advantages are obtained at a price, and that price is
particularly high when the survey geometry is highly irregular. With a
pencil beam survey, for instance, the psf is a pancake with transverse
dimension $1/w$, where $w$ is the width of the needle, and the direct
estimate of the power at low spatial frequencies will be strongly
contaminated by `aliasing' from high spatial frequencies (Broadhurst
\etal\ 1990;   Kaiser \& Peacock   1991). The interpretation   of the
results are then quite difficult and requires careful modeling of the
convolution. In a direct estimate of $\xi(r)$ no such problem arises,
although estimating the statistical error in $\xi$ is still
problematic for highly irregular geometries.

In this paper we concentrate on the development of a formalism that is
designed to construct a descriptive statistic that measures the power
spectrum of the underlying density fluctuation field assuming that the
fluctuation field is some homogeneous and statistically isotropic
random process.  We apply the formalism to surveys with a large
baseline in all three dimensions, in particular, to the flux limited
1--in--6 IRAS QDOT survey that covers 73.9\% of the sky and has 1824
galaxies with redshift that corresponds to radii of $20\,\hm1\, {\rm
Mpc} < R < 500\,\hm1\,{\rm Mpc}$ (where $h\equiv H_0/100\,{\rm
kms}^{-1}{\rm Mpc}^{-1}$, as usual) and galactic latitude
$|b|>10^\circ$.  We have used a revised version of the QDOT database
in which a redshift error that afflicted approximately 200 southern
galaxies has been corrected (courtesy of A. Lawrence).  Since the QDOT
survey is deep and covers nearly the whole sky we feel that in this
case, and in general for surveys of that type, power--spectrum
analysis would seem to be the method of choice.

The layout of the paper is as follows: In \S 2 we describe the method;
this is essentially the use the power spectrum of the weighted galaxy
counts, with the weight optimized for Gaussian fluctuations.  A new
feature of this work as compared with previous studies is the rigorous
error analysis and optimized weighting scheme.  In \S 3 we apply the
method to the QDOT 1--in--6 survey. The results which confirm the
impression from e.g. the counts--in--cells analysis that there is an
excess of large--scale power relative to the CDM predictions (if we
normalize to fluctuations at the $\sim 8 \hm1$ Mpc scales) and that
the excess appears as a `bump' in $P(k)$ around wavelengths
$\lambda\sim 100 - 150\hm1$ Mpc.  In \S 4 we compare the results with
other probes of the large--scale power spectrum, theoretical models
and analysis from other surveys. In \S 5 we examine the statistical
fluctuations in the power and compare these with the Gaussian
expectation. We conclude in \S 6.

\beginsection 2. METHOD

In \S 2.1 we construct our estimator of the power $\hat P(k)$. In \S
2.2 we calculate the variance in $\hat P(k)$ under the assumption that
the Fourier transform of the galaxies is approximately Gaussian
distributed. In \S 2.3 we find the optimum weighting scheme, and in
\S 2.4 we will convert the results to practical formulae to implement
on the computer. In \S 2.5 we present the covariance matrix.

\beginsection 2.1 The Estimator

We make the
usual assumption that the galaxies form a Poisson sample
(Peebles 1980) of the density field $1 + f(\br)$:
$$
P({\rm vol\; element\;}\delta V\;{\rm contains\; a\; galaxy})
= \delta V \nbar(\br) \bigl[1 + f(\br)\bigr]
\eqno(2.1.1)
$$
where $\nbar(\br)$ is the expected mean space density of galaxies
given the angular and luminosity selection criteria,
and we wish to estimate the power spectrum
$$
P(k) = P(\bk) \equiv \int d^3r\, \xi(\br)\, e^{i \bk\cdot \br}
\eqno(2.1.2)
$$
where $\xi(\br) = \xi(r) = \langle f(\br') f(\br'+\br)\rangle$.  Our
Fourier transform convention is $f(\bk) = \int\, d^3r\, f(\br)\,
\exp(i\bk\cdot\br)$ so $f(\br)$ $=$ $(2\pi)^{-3}$ $\int d^3k$ $ f(\bk)
\exp(-i\bk\cdot\br)$. Since we are dealing with a random field of
infinite extent, the numerical value of $f(\bk)$ is not well defined,
but if we set $f(\br) = 0$ outside of some enormous volume $V$ then
$\langle f(\bk)f^*(\bk) \rangle / V = \int d^3r\, \langle f(\br')
f(\br' + \br)\rangle \exp(i\bk\cdot\br)$ tends to a well defined limit
and it is this quantity that we call $P(k)$.

Our approach is to take the Fourier transform of the real galaxies
minus the transform of a synthetic catalogue with the same angular and
radial selection function as the real galaxies but otherwise without
structure.  We also incorporate a weight function $w(\br)$ which will
be adjusted to optimise the performance of the power--spectrum
estimator. We define the weighted galaxy fluctuation field, with a
convenient normalization, to be
$$
F(\br) \equiv {w(\br)\bigl[n_g(\br)
- \alpha n_s(\br)\bigr]
\over \left[\int d^3r\, \nbar^2(\br)\, w^2(\br) \right]^{1/2}}
\eqno(2.1.3)
$$
where $n_g(\br) = \sum_i\delta(\br - \br_i)$ with $\br_i$ being the location
of the $i^{\rm th}$ galaxy
and similarly for a synthetic catalogue which has number density $1/\alpha$
times that of the real catalogue.

Taking the Fourier transform of $F(\br)$, squaring it and taking the
expectation value we find:
$$
\langle |F(\bk)|^2 \rangle = {
\int d^3r\, \int d^3r'\, w(\br)\, w(\br')
\langle
[n_g(\br) - \alpha n_s(\br)]
[n_g(\br') - \alpha n_s(\br')]
\rangle
e^{i \bk\cdot (\br - \br')}
\over
\int d^3r\, \nbar^2(\br)\, w^2(\br)
}
\eqno(2.1.4)
$$
With the model of equation 2.1.1, the two point functions of $n_g$, $n_s$ are
$$
\eqalign{
\langle n_g(\br) n_g(\br') \rangle &=
\nbar(\br) \nbar(\br') [1 + \xi(\br - \br')] +
\nbar(\br) \delta(\br - \br')\cr
\langle n_s(\br) n_s(\br') \rangle &=
\alpha^{-2} \nbar(\br) \nbar(\br') +
\alpha^{-1} \nbar(\br) \delta(\br-\br')\cr
\langle n_g(\br) n_s(\br') \rangle &=
\alpha^{-1} \nbar(\br) \nbar(\br') \cr
}
\eqno(2.1.5)
$$
(see appendix A), so
$$
\langle |F(\bk)|^2 \rangle
= \int {d^3k' \over (2 \pi)^3} P(\bk') |G(\bk - \bk')|^2
 + (1 + \alpha)
{
\int d^3r\, \nbar(\br)\, w^2(\br) \over
\int d^3r\, \nbar^2(\br)\, w^2(\br)}
\eqno(2.1.6)
$$
where
$$
G(\bk) \equiv
{
\int d^3r\, \nbar(\br)\, w(\br)\, e^{i \bk\cdot\br}
\over
\left[\int d^3r\, \nbar^2(\br)\, w^2(\br) \right]^{1/2}
}.
\eqno(2.1.7)
$$
The content of this result is readily understood. The density field
in our data is the true infinite density field times some mask, so
in Fourier space we obtain a convolution between the transforms
of the true density field and of the mask. The fair--sample
hypothesis assumes that there are no phase correlations
between density and mask, so that the power spectrum of the
data is the true power spectrum convolved with that of the mask.
For Poisson--sampled density fields, we obtain in addition
a constant shot--noise contribution to the power. This
arises because the discrete density field has a delta function
in its correlation function. Since $\xi(0)$ is assumed to
be better behaved than this for the underlying density
field, this discreteness contribution is usually subtracted.

For the IRAS survey $G(\bk)$ is a rather compact function with width
$\sim 1 / D$, where $D$ characterizes the depth of the survey,
Provided
we restrict attention to $|\bk| \gg 1/D$, which is really just
the requirement that we have a `fair sample', and provided $P(\bk)$
is locally smooth on the same scale, then
$$
\langle |F(\bk)|^2 \rangle
\simeq P(\bk) + P_{\rm shot},
\eqno(2.1.8)
$$
so the raw power spectrum $|F(\bk)|^2$ is the true power spectrum plus
the constant shot noise component
$$
P_{\rm shot} \equiv
{
(1 + \alpha)\int d^3r\, \nbar(\br)\, w^2(\br) \over
\int d^3r\, \nbar^2(\br)\, w^2(\br)}.
\eqno(2.1.9)
$$
Our estimator
of $P(\bk)$ is just
$$
\hat P(\bk) = {|F(\bk)|^2 - P_{\rm shot} },
\eqno(2.1.10)
$$
and to obtain our final estimator of $P(k)$ we average
$\hat P(\bk)$ over a shell
in $k$-space:
$$
{\hat P}(k) \equiv {1\over V_k}
\int_{V_k} d^3k' {\hat P}(\bk'),
\eqno(2.1.11)
$$
where $V_k$ is the volume of the shell.

Equations 2.1.3, 2.1.9--11 provide our operational definition of $\hat
P(k)$. To use these we must specify the weight function $w(\br)$
which so far has been arbitrary, and we must choose some sampling grid
in $k$--space. In order to set these wisely --- and also to put error
bars on our estimate of the power --- we need to understand the
statistical fluctuations in $\hat P(\bk)$.

\beginsection 2.2 Statistical Fluctuations in the Power

{}From equation (2.1.11) the mean square fluctuation in $\hat P(k)$ is
$$
\sigma_P^2 \equiv
\left\langle \left({\hat P(k) - P(k)}\right)^2\right\rangle =
{1\over V_k^2}
\int_{V_k} d^3k
\int_{V_k} d^3k'
{\langle \delta \hat P(\bk)
\delta \hat P(\bk') \rangle}. \eqno(2.2.1)
$$ which depends on the two point function of $\delta \hat P(\bk)
\equiv \hat P(\bk) - P(k)$. Since $\hat P(\bk)$ is itself a two--point
function of $F(\br)$ this depends on the four--point function which is
poorly known. An interesting model for the two point function of
$\delta \hat P(\bk)$ is to assume that the Fourier coefficients
$F(\bk)$ are Gaussian distributed, in which case $\langle \delta \hat
P(\bk) \delta \hat P(\bk') \rangle =
|\langle F(\bk) F^*(\bk') \rangle|^2$ (see appendix B).
There are several factors which motivate
the Gaussian assumption. The Fourier
transform of $F(\br)$ at some low spatial frequency $\bk$ will receive
contributions both from real low frequency density fluctuations and
from small scale clustering and discreteness of galaxies. The latter
will tend to produce Gaussian fluctuations by virtue of the central
limit theorem, and, in the simplest inflationary scenarios at least,
the long--wavelength fluctuations will also be Gaussian distributed.
Perhaps a stronger motivation is that the Gaussian hypothesis seems to
agree well with the observations; see \S 5.

We can calculate $\langle F(\bk) F^*(\bk') \rangle$
by a simple generalisation of the steps leading to 2.1.6, and we find
$$
\langle F(\bk)F^*(\bk') \rangle
= \int {d^3k'' \over (2 \pi)^3} P(\bk'') G(\bk - \bk'') G^*(\bk' - \bk'')
 + S(\bk' - \bk),
\eqno(2.2.2)
$$
where we have defined
$$
S(\bk) \equiv {
(1+\alpha) \int d^3r\, \nbar(\br)\, w^2(\br)\, e^{i\bk\cdot\br}
\over
\int d^3r\, \nbar^2(\br)\, w^2(\br),
}
\eqno(2.2.3)
$$
and, in the same approximation that led to equation (2.1.8) we obtain
$$
\langle F(\bk)F^*(\bk + \delta \bk) \rangle
\simeq P(\bk) Q(\delta \bk) + S(\delta \bk)
\eqno(2.2.4)
$$
where
$$
Q(\bk) \equiv
{
\int d^3r\, \nbar^2(\br)\, w^2(\br)\, e^{i\bk\cdot\br}
\over
\int d^3r\, \nbar^2(\br)\, w^2(\br)
},
\eqno(2.2.5)
$$
and therefore
$$
\langle \delta \hat P(\bk) \delta \hat P(\bk') \rangle
= |P(\bk) Q(\delta \bk) + S(\delta \bk)|^2.
\eqno(2.2.6)
$$

Under the Gaussian assumption, $F(\bk)$ and $\delta \hat P(\bk)$ take
the form of locally homogeneous random processes with 2--point
functions whose shapes are determined by the survey geometry.  The
variance in the power $\langle \delta \hat P(\bk)^2 \rangle$ is just
the square of the total power (i.e. signal power plus shot noise
power), and the two point function of the power is again a compact
function with width $\sim 1/D$. Thus, the estimator of the power
behaves like an incoherent random field smoothed on scale $\delta k
\sim 1 /D$.  Note the relation between $Q(\delta\bk)$ and
$G(\delta\bk)$ defined in 2.1.7; these are both measures of the
coherence in $k$ space.  The fact that $G$ depends on $\bar n$ and $Q$
on ${\bar n}^2$ is just the usual difference between the transform of
a field and the transform of its two--point function.

In this regard the power spectrum estimator is very different from the
correlation function estimator. If we have some continuous field
$f(r)$ that we view through a survey `window' of scale $D$, then as
$D$ becomes large the estimator of the two--point function tends
asymptotically to $\xi(r)$. The microscopic fluctuations in $\hat
P(\bk)$, in contrast, remain of order unity even as $D\rightarrow
\infty$, but the coherence length for the fluctuations shrinks so when
we average over some finite volume of frequency space the averaged
power tends to $P(k)$, and the fractional fluctuations in $\hat P(k)$
will be on the order of the square root of the number of coherence
volumes averaged over in equation 2.1.11.

Equation 2.2.1, 2.2.6 can be used in a number of ways: With $P(k) =
\hat P(k)$ we obtain self--consistent error bars;
with $P(k) = P_{\rm\scriptscriptstyle CDM}(k)$, for instance,
we obtain the expected fluctuations for this
specific theory, and with $P(k) = 0$ we obtain the `Poisson' error
bars widely used in the past, though these of course tend to
underestimate the real uncertainty.

It should be stressed that the Gaussian model is only an assumption.
An alternative would be to estimate $\langle \delta \hat P(\bk) \delta \hat
P(\bk + \delta \bk) \rangle$ directly from $\hat P(\bk)$ --- as well
as providing an empirical error estimate, by comparing this directly
with equation 2.2.6 we could obtain an interesting test of the Gaussian
nature of the fluctuations. This issue will be discussed in greater depth
in \S 5.

\beginsection 2.3 Optimum Weighting

If the shell we average over in equation 2.1.11 has a width which is large
compared to the coherence length then the double integral in
2.2.1 reduces to
$$
\sigma_P^2(k) \simeq {1\over V_k}
\int d^3k\,' |P(k)Q(\bk') + S(\bk')|^2,
\eqno(2.3.1)
$$
so, with the definition of $Q(\bk)$ and $S(\bk)$ and using Parseval's theorem,
the fractional variance in the power is
$$
{\sigma_P^2(k) \over P^2(k)}
=
{(2\pi)^3\int d^3r\, \nbar^4\, w^4 [1 + 1/\nbar P(k)]^2 \over
V_k \left[ \int d^3r\, \nbar^2\, w^2 \right]^2}.
\eqno(2.3.2)
$$

We seek $w(\br)$ which minimises this. Writing $w(\br) = w_0(\br)
+ \delta w(\br)$ and requiring that $\sigma_P^2(k)$ be stationary
with respect to arbitrary variations $\delta w(\br)$ we obtain
$$
{
\int d^3r\, \nbar^4\, w_0^3 \left({1 + \nbar P \over \nbar P}\right)^2 \delta
w(\br)
\over
\int d^3r\, \nbar^4\, w_0^4 \left({1 + \nbar P \over \nbar P}\right)^2
} =
{
\int d^3r\, \nbar^2\, w_0 \delta w(\br)
\over
\int d^3r\, \nbar^2\, w_0^2
}
\eqno(2.3.4)
$$
and it is easy to see by direct substitution that this is satisfied if we
take
$$
w_0(\br) = {1 \over 1 + \nbar(\br) P(k)}.
\eqno(2.3.5)
$$
This is the optimal weighting (under the assumption that the fluctuations
are Gaussian). It is the analogue of the ``$1 + 4 \pi \nbar J_3$''
weighting scheme often used in correlation analysis, and, just as in that
case, the procedure is somewhat circular since one needs a preliminary
estimate of $P(k)$ in order to set the weighting.

The asymptotic behavior of this weighting scheme is very reasonable:
Say we want to measure density fluctuations on a particular scale
$\lambda$. Since $\nbar(r)$ is a rapidly decreasing function we
will have two regimes: For small $r$, we will have many galaxies per
$\lambda^3$ volume, so the error will be dominated by the finite
number of independent `fluctuation volumes' and consequently one would
like to give equal weight per volume, or equivalently to weight
galaxies in proportion to $1/\nbar$. At large radius we are dominated
by shot--noise and consequently one would like to weight galaxies
equally. Equation 2.3.5 is in accord with these notions.
Note that the optimal weight depends on the spatial frequency. Insofar as the
power tends to decrease with frequency this means that when we measure
long wavelengths we will be inclined to give greater weight to the more
distant galaxies.

An important exception to the above analysis arises when the
sampling of the density field is not Poissonian. This arises
in practice when the sky is divided into a number of zones,
and a fixed number of redshifts per zone are measured,
independent of the actual number of galaxies present
(e.g. the Campanas redshift survey: Shectman \etal\ 1992).
Clearly, a gross underestimate of the power spectrum would
result if we were simply to treat the measured redshift
data as a Poisson sample. The correct procedure is to weight
the galaxies according to the sampling factor: if a given
galaxy is part of the fraction $f$ sampled from a given field,
then we form the density field $n_g({\bf r})=\sum_i f_i^{-1}
\delta ({\bf r-r_i})$. This gives an estimate of the
density field which is corrected for the variable sampling,
and all that alters is the shot noise term. The first of equations
2.1.5 now becomes
$$
\langle n_g(\br) n_g(\br') \rangle =
\nbar(\br) \nbar(\br') [1 + \xi(\br - \br')] +
\nbar(\br)f^{-1}(\br) \delta(\br - \br')
\eqno(2.2.7)
$$
so that the shot term in 2.1.6 is greater then it would have
been for the same number of galaxies with constant sampling.
Otherwise, the analysis goes through as before.
This may seem a little surprising, given that the
sampling factors are not imposed in advance, but `know'
about the large--scale properties of the density field.
However, study of the derivation of equation 2.1.5 given in
appendix A shows that there is no problem. The critical
term is the evaluation of $\langle n_i n_j\rangle$ for
two different cells. Even if the mean density of objects
has been adjusted to reflect the average density perturbation
over some region, galaxies are still distributed at random {\it within\/}
that region, which is all that is required in order to
write $\langle n_i n_j\rangle$ in terms of $\xi({\bf r_i-r_j})$.

\beginsection 2.4 A Practical Algorithm.

We now summarize the results, and write all the integrals in terms of
discrete sums as will be evaluated on the computer. Let us assume
that we are provided with the coordinates for the real and synthetic
catalogues plus a function which returns the value of $\nbar(\br)$
(the details of how the synthetic catalogue was actually constructed
for the IRAS survey are given below). Alternatively, the local value
of $\nbar$ may be provided along with the coordinates of the synthetic
galaxies.

We need to evaluate $F(\bk)$ and also the auxiliary functions
$Q(\bk)$ and $S(\bk)$. We know that these will be smooth on scale
$\delta k \sim 1/ D$, so we evaluate these on a cartesian grid. We
would like to emphasize that the geometry of the survey requires
a cartesian grid, and so we sample $k-$space linearly and present the
results as linear--log plots rather than the traditional log--log
plots. A reasonable value for the grid spacing for the IRAS survey is
$\delta k = 0.02 h$ Mpc$^{-1}$. Since the functions $Q(\bk)$ and
$S(\bk)$ will be rather compact we need only evaluate these on a
fairly small grid. A convenient way to perform the spatial integrals
is to use $\int d^3 r\,\nbar(\br) \ldots \rightarrow \alpha \sum_s
\ldots$, where the sum is over the synthetic galaxies which we assume
are sufficiently numerous to define $\nbar$. In order to set the
weight scheme (equation 2.3.5) we need to assume some value for
$P(k)$. The approach we have taken is to use a range of values for
$P(k)$ --- each of which provides a legitimate though not necessarily
optimal estimate --- and then one can select, for any range of
wavenumber, the appropriate optimal estimator. Having chosen a value
for $P(k)$ it is convenient to adjust the normalization of the weight
function so that
$$
\int d^3r\, \nbar^2(\br)\, w^2(\br) \rightarrow \alpha \sum_s \nbar(\br_s)\,
w^2
(\br_s) = 1
\eqno(2.
4.1)
$$
we then evaluate
$$
F(\bk) = \int d^3r\,\, w(\br)[n_g(\br) - \alpha n_s(\br)] e^{i\bk\cdot\br}
\rightarrow
\sum_g w(\br_g)e^{i\bk\cdot\br_g} - \alpha \sum_s w(\br_s)e^{i\bk\cdot\br_s},
\eqno(2.4.2)
$$
$$
Q(\bk) = \int d^3r\, \nbar^2(\br)\, w^2(\br)\, e^{i\bk\cdot\br}
\rightarrow
\alpha \sum_s \nbar(\br_s)\, w^2(\br_s) e^{i\bk\cdot\br_s}
\eqno(2.4.3)
$$
and
$$
S(\bk) = (1 + \alpha) \int d^3r\, \nbar(\br)\, w^2(\br)\, e^{i\bk\cdot\br}
\rightarrow
\alpha (1 + \alpha) \sum_s w^2(\br_s) e^{i\bk\cdot\br_s}.
\eqno(2.4.4)
$$
Our radially averaged power spectrum estimator is
$$
\hat P(k) = {1\over N_k} \sum\limits_{k < |\bk| < k + \delta k}
|F(\bk)|^2 - S(0)
\eqno(2.4.5)
$$
where $N_k$ is the number of modes in the shell.

In \S 2.3 we obtained the rather simple expressions 2.3.1, 2.3.2 for the
variance in the power. These say that the rms fractional fluctuation in
the power is just the square root of the number of `coherence volumes' in
$k$--space: $\sigma_P(k) / P(k) = \sqrt{V_c / V_k}$, where the coherence
volume is defined by
$V_c \equiv (2\pi)^3 \int d^3r\, \nbar^4\, w^4 (1 + 1/ \nbar
P)$. However, these require that the shell be thick compared to the coherence
length. This is fine at high frequencies $k \gg 1 / D$, but at low
frequency this would result in a loss of resolution.
Instead, we use the analogue of 2.2.1
$$
\sigma_P^2(k) = {2\over N_k^2}
\sum_{\bk'}
\sum_{\bk''}
|P Q(\bk' - \bk'') + S(\bk' - \bk'')|^2
\eqno(2.4.6)
$$
where $\bk$ and $\bk'$ are constrained to lie in the shell and
which is valid for any shell thickness.
If we make the shells thinner than or comparable to the coherence length
then neighboring shells will be correlated, but the variance tends to
a well defined finite value even as the shell tends to zero width.

\beginsection 2.5 Covariance Matrix

The error estimate 2.4.6 is adequate to give an estimate of the
variance in the power spectrum, but for a more precise assessment
of theoretical models it is necessary to quantify the degree of
correlation of $\hat P(k)$ at differing wave numbers. The way to
do this is via likelihood analysis.

Provided the fractional error is moderately small (i.e. the shell
intercepts a sufficiently large number of coherence volumes), the
fluctuations in the power will themselves tend to become Gaussian
distributed, and the vector of estimates $\hat P_i$ together with the
correlation matrix allow one to evaluate the likelihood for any
particular theory:
$$
L[P_{\rm th}(k)] = p[P_i | P_{\rm th}(k)] =
	{e^{- C_{ij}^{-1} [\hat P_i - P_{\rm th}(k_i)]
	[\hat P_j - P_{\rm th}(k_j)]/2}
	\over
	(2\pi)^{N/2} |c|}\ .
\eqno(2.5.1)
$$
This provides a quantitative way to compare theories.

This is rather similar to other applications of likelihood analysis to
cosmological data sets (Kaiser 1992, Bond \& Efstathiou 1984), but
there is a slight difference here in that the correlation matrix for
the binned estimates of $\hat P$ actually depends on $P(k)$ itself:
$$
C_{ij} \equiv \langle \delta\hat P(k_i) \delta\hat P(k_j)\rangle =
 	{2 \over N_k N_{k'}}
	\sum\limits_{\bk}\sum\limits_{\bk'}
		|P Q(\bk - \bk') + S(\bk - \bk')|^2
\eqno(2.5.2)
$$
where $\bk$ and $\bk'$ lie in the shells around $k_i$ and $k_j$
respectively.

Equation 2.5.2 is derived, like equation 2.2.6, under the assumption that
$P(k)$ is effectively constant over a coherence scale. With this
assumption we can write
$$
C_{ij} = C_{ij}^{(0)} + C_{ij}^{(1)} \sqrt{P_{\rm th}(k_i)P_{\rm th}(k_j)}
+ C_{ij}^{(2)}P_{\rm th}(k_i)P_{\rm th}(k_j)
\eqno(2.5.3)
$$
The use of this will be illustrated in \S 4.
\vfill\eject
\beginsection 3. THE QDOT SURVEY

\beginsection 3.1 The Survey.

We apply the analysis to the updated (A. Lawrence 1993, private
communication) IRAS QDOT survey (Efstathiou \etal\ 1990).  The QDOT
survey chooses at random one in six galaxies from the IRAS point
source catalogue with a flux limit at $60\mu$m $f_{60} > 0.6$Jy.  In
this sample there are 1824 galaxies above galactic latitude
$|b|>10^\circ$ with redshifts that corresponds to radii
$20\,\hm1\,{\rm Mpc}<R<500\,\hm1\,{\rm Mpc}$.  We have used a revised
version of the QDOT database in which a redshift error that afflicted
approximately 200 southern galaxies has been corrected.  We have
converted all redshifts to the local group frame.

The survey provides us with an angular mask. The sky is divided into
41167 bins, each $\approx 1^\circ \times 1^\circ$ , some of which are
masked. Since most of the bins below galactic latitude $|b|<10^\circ$
are masked because of obscuration due to the galactic plane, we masked
off all bins that had $|b|<10^\circ$. There are 30444 unmasked bins, a
coverage of $\approx74\%$ of the sky. The QDOT galaxy distribution and
mask are shown in figure 1. A more complete description of the sample
is given in Efstathiou \etal\ (1990).

To apply the above formalism to the QDOT survey, we wrote two
independent codes that utilized different grids in $k$--space and
different forms for the radial selection function. We obtained very
good agreement between the results for all values of $P(k)$ in the
weight function (Eqn. 2.3.5).

\beginsection 3.2 Construction of the Synthetic Catalogue

Once we have the radial galaxy distribution from the survey, we bin the
galaxies in bins of width $dr$ and then fit, using a $\chi^2$
technique, a number function
$$
n(r) = A \left({r\over r_{\rm max}}\right)^x \left[
		1 + \left({r\over r_{\rm max}}\right)^y
	\right]^{-{y+x\over y}}
\eqno(3.2.1)
$$
where
$$
A = 2^{y+x\over y} n_{\rm max}\ ,
\eqno(3.2.2)
$$ $n_{\rm max}\equiv n(r=r_{\rm max})$ and $r_{\rm max}$ is the
radius where $n(r)$ peaks (see figure 2 for the radial distribution of
the IRAS QDOT galaxies and a best fit). We integrated $n(r)$ and
divided the function into $n_{_n}$ bins, each of which has the same
number of galaxies.  To check the dependence on the exact choice of
the radial selection function, we also tried log--normal bins and
varied the parameters in the fitting function Eqn. 3.2.1 to give us a
$1\sigma$ effect. The effect of these choices on the final results is
negligible. We constructed the synthetic catalogue by distributing
galaxies with a Poisson distribution in all unmasked bins given the
above selection function.

\beginsection 3.3 The Power spectrum.

We have seen that the optimum weight function depends on $k$
(Eqn. 2.3.5), so that a different weighting should in principle
be used at each wavenumber. This would be rather cumbersome,
and so we have chosen in practice to evaluate the weight
assuming a given constant level of power (\ie a $n=0$ white
noise power spectrum). By varying this assumed power over values that
cover the observed power, it is easy to see what effect the
exact choice of weight would have. We shall consider four
different assumed power levels, which produce the
different effective survey depths shown in figure 3.
In figure 4 we show the power spectrum results for the full IRAS
QDOT survey using these different weight function parameterizations. We see
that the larger the power in the weight function is (\ie the greater
the effective depth of the survey) the more power we get.
The effect is most marked at the shallow end: we gain roughly
a factor 1.5 in power when the assumed power changes from
$2000\,(h^{-1}{\rm Mpc})^3$ to $4000\,(h^{-1}{\rm Mpc})^3$,
but things change relatively little thereafter. This suggests
that the effect is a local one that represents true sampling
fluctuations, but that the correct average power is detected
for greater depths. We shall generally adopt the results with
$P=8000\,(h^{-1}{\rm Mpc})^3$ as our `standard' set of values,
since this power is closest to the average level in the $k$
range of interest.

We attempted to choose different selection criteria that might help us
disentangle clustering effects that arise because the more distant
galaxies are intrinsically more luminous than those which are nearby.
We divided the data into six samples with different flux limits, and
for each we chose a number of weight functions by changing the power
$P(k)$ in $w(\br)$. We found no systematic effects that appeared to
depend primarily on luminosity, rather than distance.

The most interesting feature of the power--spectrum results is that the
function is relatively smooth, with no significant sharp features.  As
is discussed in more detail below, there is an excess at $0.03<k<0.07$
when comparing to the CDM power spectrum with $\Omega h=0.5$,
normalized around $k=0.1$.  This occurs for all weight functions we
used as well as for most sub--samples.  At the largest wavelengths,
our data indicate a turnover in the power spectrum, with reduced power
for $k\lsim0.04$ corresponding to $\lambda\gsim150$ h$^{-1}$ Mpc.
Although the error bars are large, this also is a feature which
appears to be robust with respect to changing the depth of the survey.
There is the worry that the turnover may be a normalization effect:
our power spectra must vanish at $k=0$ since we obtain the mean
density from the data.  Whether this is a problem depends on the
convolution function $G(k)$ (Eqn. 2.1.7), and is an issue discussed by
Peacock \& Nicholson (1991). If we approximate the function by a
Gaussian $|G(k)|^2=\exp-k^2 R^2$, then the power--spectrum estimate is
sensitive to normalization effects only for $k\lsim1.5/R$.  The
appropriate values of $R$ that fit our $G$ function vary from $R=65\,
h^{-1}{\rm Mpc}$ for $P=2000\,(h^{-1}{\rm Mpc})^3$ to $R=105\,
h^{-1}{\rm Mpc}$ for $P=16000\,(h^{-1}{\rm Mpc})^3$.  The turnover in
the QDOT power spectrum is thus not an artifact of self--normalization.
Furthermore, the location of the turnover agrees well with the
position of the same feature as seen in other data sets (e.g.  Peacock
\& West 1992).

\beginsection 4. DISCUSSION OF POWER--SPECTRUM RESULTS

In \S 4.1 we compare our results to other probes and in \S 4.2 we present
theoretical model power spectra and compare them to the QDOT results. In
\S 4.3 we present the results of CDM likelihood analysis in redshift
space and in \S 4.4 we transform the results to real space.

\beginsection 4.1 Comparison With Other Probes

In figure 5 we show a comparison between the shallow [$P(k)=2000$
($\hm1$ Mpc)$^3$] IRAS QDOT power spectrum and the 1.2--Jy Berkeley
survey (Fisher \etal\ 1993). The error bars on the 1.2--Jy survey are
derived from the standard deviation of 10 `mock' IRAS standard CDM
simulations (see Fisher \etal\ 1993 for details).  The analysis method
used on the 1.2--Jy data differs somewhat from that used here; they
weighted volumes equally within some cylindrical volume of varying
size, up to a maximum of length $180 h^{-1}$ Mpc by radius $90 h^{-1}$
Mpc.  This is equivalent in volume to a sphere of radius $103h^{-1}$
Mpc.  Comparing with figure 3, we see that our shallowest weight does
give roughly constant weighting to volumes out to this radius, which
is why we have presented the comparison in this way.

Once scaled to the same depth, the surveys agree reasonably well. The
1.2--Jy data lie slightly below our shallow results; however, there
would be a much more marked discrepancy if we had chosen to compare to
the `standard' $P=8000 (\hm1{\rm Mpc})^3$ QDOT results --- about a
factor of 1.5 in power. Also, because of their choice of sampling in
$k-$space, the 1.2--Jy resolution is not very high for large scales
(small $k$) which perhaps led them to miss some of the structure of
the power at scales $\approx 100--150\,\hm1$ Mpc.  As mentioned above,
the survey geometry requires linear sampling of $k-$space, and so
linear plotting of results.

In figure 6 we show a comparison with data from the CfA surveys (Vogeley
\etal\ 1992). The power spectra are of galaxies from the CfA 1 and the
CfA 2 surveys, divided into two categories: CfA 100 and the deeper CfA
145 (for details see Vogeley \etal\ 1992). The error bars are derived
from estimating the 95\% confidence level from the variation of 100
`mock' CfA surveys from open CDM simulations.  As with QDOT, the CfA
results appear to show some weak trend for the power to increase with
increasing sample depth.  The shapes of the CfA and QDOT power spectra
appear to differ: similar levels of power are seen at large
wavelength, but the CfA spectra show much more small--scale power.
However, note the relative sizes of the large---wavelength error bars in
the CfA data, plus the fact that their analysis gives no idea of the
extent of any cross--correlation between different points. At present,
it is probably not possible to rule out with great confidence the
hypothesis that CfA and QDOT power spectra have the same shape, but an
amplitude different by a factor of about 2. Improved data will be
interesting here, as a difference in the shapes of the power spectra
for optically--selected and IRAS galaxies must hold information about
the processes which created the differing spatial distributions for
these objects.

\beginsection 4.2 Fitting Model Power Spectra

It is interesting to fit the power--spectrum data by some analytical
model. Since the result appears to be a relatively smooth function, it
is convenient to use for this purpose the CDM linear power spectrum.
This allows for parameterization of the slope and amount of curvature,
depending on the density. Initially, we shall use this simply as an
empirical way of describing the data; the physical interpretation of
the results will be given later.

The CDM power spectrum takes the form of a scale--invariant spectrum
modified by some transfer function: $P(k)\propto k\, T_k^2(k)$. We
shall use the BBKS approximation (Bardeen \etal\ 1986), which is the
most accurate fitting formula:
$$
T_k={\ln(1+2.34q)\over 2.34q}[
	1+3.89q+(16.1q)^2 +(5.46q)^3 +(6.71q)^4
	]^{-1/4},
\eqno(4.2.1)
$$
where $q\equiv k/[\Omega h^2\; {\rm Mpc}^{-1}]$. Since observable
wavenumbers are in units of $h\,{\rm Mpc}^{-1}$, the shape parameter
is the apparent value of $\Omega h$. [This should not be confused with
the parameter $\Gamma$ defined by Efstathiou, Bond \& White (1992);
they considered a transfer function which fits a CDM model with
non--zero baryon density $\Omega_B=0.03$. Empirically, the wavenumber
scaling in the CDM model depends very nearly on $\Omega
h^2\exp[-2\Omega_B]$; Efstathiou {\it et al.} defined $\Gamma=0.5$ to
correspond to $\Omega h=0.5$, and our values of $\Omega h$ are thus
smaller than $\Gamma$ by a factor of 0.94.] The normalization of the
power spectrum can be expressed in various ways (e.g. Peacock 1991);
here we shall use $\sigma_8$ --- the linear rms in spheres of radius
$8h^{-1}$ Mpc.

In figure 7 we show a comparison with a `standard' linear CDM power
spectrum having $\Omega h=0.5$. The poor fit is apparent, although
whether one regards this as the data having an excess of large--scale
power or a lack of small--scale power is a matter both of taste and of
where we choose to normalize.  We further compare the results to the
APM fitting formula of Peacock (1991):
$$
P(k) = k^{-3}{(k/k_0)^{1.6}\over 1 + (k/k_c)^{-2.4}}\ .
\eqno(4.2.2)
$$
A reasonable fit to the IRAS QDOT power spectrum is achieved with
$k_0 = 0.03\, h\, {\rm Mpc}^{-1}$, and $k_c \in [0.025,0.04]\, h\,
{\rm Mpc}^{-1}$.
The overall shape and position of the break agree
reasonably well with the APM data; however, note that $k_c\simeq 0.015
h\, {\rm Mpc}^{-1}$ is required to best--fit the APM data, and
such a large break wavelength appears to be excluded by our data.

We also compare our results with simulations of mixed [cold (70\%)
plus hot (30\%)] dark matter (MDM) (Klypin \etal\ 1992).  The `red
galaxies' power spectrum in the MDM simulations is the one of all the
dark matter halos. The halos were defined to be at the maxima of the
overdense regions with overdensity $>50$. The `galaxies' were
displaced along the line of sight in accordance with there peculiar
velocities to mimic redshift space and the density field was smoothed
with a Gaussian filter of $1/2$ a cell size radius to reduce shot
noise (Klypin 1993, private communication). The overall shape of the
power of the red galaxies in the MDM simulations agrees quite well
with our power spectrum.  As we shall see below, this is because the
MDM scenario gives results that are rather similar to those of CDM
models with low $\Omega h$ or `tilted' spectra.

Rather than looking at {\it a priori\/} models any further, we
shall now proceed to fit linear CDM power spectra to the QDOT
redshift--space power--spectrum data. We emphasize that the resulting
values of shape ($\Omega h$) and normalization ($\sigma_8$) {\it are
apparent redshift--space values only}. Before they can be interpreted,
we shall need to consider the effects of nonlinearities and distortions
caused by the mapping between real space and redshift space.

\beginsection 4.3 Results in Redshift Space

Since we have a procedure for constructing the power spectrum
covariance matrix for any given model (equation 2.5.3), it is easy to fit
the CDM models correctly to our data using maximum likelihood. To
within a constant, the likelihood is
$$
-\ln L = \chi^2/2 +[\ln {\rm det}\, C]/2,
\eqno(4.3.1)
$$
and
$$
\chi^2=C^{-1}_{ij}[\hat P-P_{\rm\scriptscriptstyle CDM}]_i
 [\hat P-P_{\rm\scriptscriptstyle CDM}]_j,
\eqno(4.3.2)
$$
where $C_{ij}$ is the covariance matrix for our data.

Figure 8 shows contours of likelihood at $-\ln L = {\rm minimum} +
0.5$, 1, 2, \dots. On a Gaussian approximation, the 95\% confidence
level would be at $\Delta\ln L=3$. This plot shows that the
maximum--likelihood model is well defined (and is an excellent
description of the data: $\chi^2=12$ on 18 degrees of freedom). The
best--fitting parameters and their rms uncertainties are
$$
\eqalign{
\sigma_8 &= 0.88 \pm 0.07 \quad ({\rm redshift\ space}) \cr
\Omega h &= 0.25 \pm 0.08 \quad ({\rm redshift\ space}) \cr
}
\eqno(4.3.3)
$$

We may also consider the use of the CDM transfer function with a power
spectrum which is not scale--invariant. Possibilities include `tilted'
models:
$$
P\propto k^n;\ n\simeq 0.8
\eqno(4.3.4)
$$
(Cen {\it et al.} 1992) or an inflationary prediction for logarithmic
corrections to a scale--invariant spectrum (Kofman, Gnedin \& Bahcall
1993). The latter corrections have a negligible effect on our results,
but the use of tilted models is important. The apparent value of
$\sigma_8$ is insensitive to the assumed value of $n$, but the
apparent redshift--space value of $\Omega h$ changes approximately as
$$
\Omega h = 0.25 + 0.29([1/n]-1).
\eqno(4.3.5)
$$

Comparing these results with those from the Berkeley 1.2--Jy survey
(Fisher {\it et al.} 1992), we find reasonable agreement. Their
apparent value of $\sigma_8$ is 0.80, with a best--fitting $\Omega
h=0.2$. These figures are well within our confidence limits.
This is perhaps a little surprising, since we have seen earlier
that there appears to be a difference in power of about a factor 1.5
between the 1.2--Jy results and the deeper QDOT results which are
used here. On this basis, we would have predicted a lower
value of $\sigma_8$ for the Berkeley data; the reason for
this discrepancy is unclear.

\beginsection 4.4 Results in real space

Our observed power spectrum in redshift space differs in three ways
from the quantity of interest, which is the underlying linear--theory
power spectrum of mass fluctuations. This is altered by nonlinear
evolution, by redshift--space mapping and by bias.

The last of these is easily dealt with through ignorance: we shall
assume scale--independent bias relating the {\it nonlinear\/} power spectra
$$
P_{\rm real}=b^2 P_{\rm mass}.
\eqno(4.4.1)
$$
It is clearly reasonable to treat $b$ as a constant if the galaxy
distribution is close to being unbiased. For a significant degree of
bias, one really needs a model; empirically, if our conclusions
conflict with other data, this could be interpreted as saying that $b$
must depend on scale.

The mapping between real and redshift space
introduces two effects. On large scales, there is
the linear increase of power described by Kaiser (1987):
$$
P(k)\rightarrow P(k)\left[ 1 + {2\Omega^{0.6}\over 3b}
+ {\Omega^{1.2}\over 5b^2}\right].
\eqno(4.4.2)
$$
On small scales, there is the filtering effect of virialized
peculiar velocities to deal with; measured redshifts will also be of
limited precision. These effects can be treated in the same way:
consider the simplest possible case in which these effective errors
constitute a Gaussian scatter characterized by some spatial rms
$\sigma$. In azimuthal average, this gives
$$
P(k)\rightarrow P(k)\, {\sqrt{\pi}\over2}\, {{\rm erf}\,(k\sigma)\over
 k\sigma}
\eqno(4.4.3)
$$
(Peacock 1992); modes at high $k$ are thus only damped by one power of
$k$, rather than exponentially. Fisher {\it et al.} (1992) show that
the combination of these factors describes the relation between power
spectra in real and redshift space quite accurately. Small--scale
velocities of about 200 kms$^{-1}$ rms are observationally
appropriate, and QDOT redshifts have a typical error of 300
kms$^{-1}$, making a total effective $\sigma$ of $4.4h^{-1}{\rm Mpc}$.
At the largest wavenumber considered here ($k=0.2h\, {\rm Mpc}^{-1}$),
the corresponding correction factor to the power is 0.85. This latter
correction is unavoidable, since it is based largely on uncertainties
in the data. This effect alone alters the best--fitting values of the
spectral parameters to $\sigma_8 = 0.92 \pm 0.07$ and $\Omega h = 0.28
\pm 0.08$.

The nonlinear distortions of the spectrum can be dealt with
analytically (assuming $\Omega=1$) by using the remarkable formulae
given by Hamilton {\it et al.} (1991). The result for CDM--like spectra
is that power is removed from wavenumbers $k\simeq 0.1h\,{\rm
Mpc}^{-1}$, which lowers the apparent value of $\Omega h$.

The result of applying all these corrections can be expressed in
terms of the change in the best--fitting values of
$\Omega h$ and $\sigma_8$ as a function of $b$:
$$
\eqalign{
b\,\sigma_8 &\simeq 0.92 -0.18/b^{0.8} \quad ({\rm recovered\ linear\ value})
\cr
\Omega h &\simeq 0.28 +0.05/b^{1.3} \quad ({\rm recovered\ linear\ value}) \cr
}
\eqno(4.4.4)
$$
Consistency with the values $\sigma_8=0.57$ deduced by White,
Efstathiou \& Frenk (1993) requires $b=1.37$ and $\Omega h=0.31\pm
0.08$. This last figure is in remarkable agreement with the $\Omega
h=0.32 \pm 0.07$ deduced from the cluster--galaxy cross--correlation
function by Mo, Peacock \& Xia (1993).

If the density parameter is low, the linear amplification of power in
redshift space does not occur, so that the nonlinear real--space value of
$b\,\sigma_8$ will be higher: close to the observed redshift--space
value of 0.92. The formulae of Hamilton {\it et al.} (1991) do not
apply to the low--density case, so we cannot say so exactly what the
effects of nonlinearities will be. However, experience with $N$--body
simulations suggest that the recovered linear values of
$\sigma_8$ and $\Omega h$ will still be altered
by an amount of the order 0.1 (for $b=1$) --- as in
the $\Omega=1$ case.

\beginsection 5. TESTS FOR NON--GAUSSIANITY

The expectation value of the power that we have tried to estimate
contains in itself no information about phase correlations or higher
than two--point correlations. As we have seen however, the estimated
power will inevitably fluctuate strongly from point to point in
$k$--space, and these fluctuations about the mean power are
rich in information about higher--order correlations.

One probe of non--Gaussianity is from the 1--point (in $k$--space that
is) distribution of the power. According to the Gaussian hypothesis,
the power should be exponentially distributed:
$$
p(>\hat P) = \exp(-\hat P / \overline P)
\eqno(5.1)
$$
(Kendall \& Stewart 1977) and this has been exploited as a way to
quantify evidence for periodicity in pencil--beam surveys for instance
(Szalay \etal\ 1990). The pencil--beam analysis did indeed appear to
show some evidence for non--Gaussianity: the distribution of the power
was enhanced at high values of $\hat P$. However, it can plausibly be
argued that what is happening here is that the distribution of power
was calculated over a wide range of wave--numbers and that the high
frequency components were suppressed by smoothing by redshift errors
and by random motions (Kaiser \& Peacock 1991).  Thus what one sees in
the pencil--beam case is a blend of exponential distributions with
different length scales. One way to resolve this would be to restrict
the range to low frequencies only, but unfortunately then the number
of independent `coherence cells' is rather small (even though the
baseline for this sample is impressively long).

Because we work in three dimensions, the present work yields a
much larger number of independent estimates of the
low--frequency power. The distribution of the power for $k<0.1$
(wavelengths $> 60h^{-1}$ Mpc) is shown in figure 9. The agreement with
the exponential prediction is remarkably good. It should be kept in
mind that what we are seeing here is the distribution of the `raw'
power which contains a superposition of the shot noise power, whose
long wavelength Fourier components is essentially guaranteed to be
Gaussian distributed by virtue of the central limit theorem. For the
wavelengths employed in figure 9, the real long--wavelength power is
roughly equal to the shot noise power.
The fact that the exponential law is nevertheless exact out to
$\hat P/\bar P\simeq 10$ is thus an extremely exacting test of
the Gaussian hypothesis --- and should provide an important constraint
on non--Gaussian models such as those based on topological defects.

There are potentially many other statistics that one could construct
based on the Fourier components which would measure interesting high
order correlations. One particularly interesting one is the two--point
function of the power $\chi(\delta k; k) \equiv \langle \hat P(k) \hat
P(k + \delta k)\rangle$. As we saw in \S 2.4, if we assume Gaussian
fluctuations then $\chi(\delta k; k)$ is simply determined from the
geometry of the survey (it is essentially the Fourier transform of the
survey volume) and will therefore have width in $\delta K$ on the
order of $1/D$. It is not difficult to see however how this
prediction might be modified for certain rather interesting
non--Gaussian models. Consider, for instance, a density field which is
the product of a Gaussian field with a `modulating' field which has
only long wavelength components. In such a model, the density field
would look locally Gaussian, but seen on a larger scale one would have
patches of greater or lesser amplitude. The micro--scale fluctuations
in $\hat P(k)$ in such a model would be the same as in a pure Gaussian
model but with a patchy survey volume; i.e.\ the two point function
would be more extended than predicted by equation (2.4.6). The excess
width of the $\chi$ statistic therefore measures the degree to which
fluctuations of spatial frequency $k$ are being modulated by
frequencies $\sim \delta k$.

\beginsection 6. CONCLUSIONS

We have presented a formalism for power--spectrum analysis of fully
three--dimensional deep redshift surveys.  Our main new result is an
analytical estimation of the statistical uncertainties in the power
(including both sampling and galaxy counting statistics). We have also
presented a rigorous analytical formulation of the optimal weight
function for the data, assuming that the long--wavelength Fourier
components are Gaussian distributed.  Assuming that the power is
smooth, we have shown how to derive the full covariance matrix for the
power spectrum.  This provides all the information necessary for a
proper statistical comparison between power--spectrum data and theory.

We have applied the method to the updated 1--in--6 QDOT IRAS redshift
survey.  We find that survey depths in excess of $100h^{-1}$ Mpc are
necessary in order to obtain a stable estimate of the power spectrum.
Our results strengthen and quantify the impression that there is extra
power on large scales as compared to the standard CDM model with
$\Omega h\simeq 0.5$.  Nevertheless, there appears to be a break in
the power spectrum at wavelengths $\lambda\approx150-200\hm1$ Mpc, with
sharply reduced power for larger wavelengths. This is consistent with
the picture emerging from a number of other studies.

We have applied likelihood analysis using the BBKS approximation of
the CDM spectrum with $\Omega h$ as a free parameter as a
phenomenological family of models: in redshift space the best--fitting
parameters are $\sigma_8=0.88\pm0.07$, $\Omega h=0.25\pm0.08$. We have
attempted to treat the distortions to the power spectrum introduced by
nonlinear evolution and the redshift--space mapping, and so recover the
parameters which describe the linear power spectrum.  If the linear
rms variance is taken to agree with White \etal\ (1993)
($\sigma_8=0.57$), we find that a linear power spectrum with $\Omega
h=0.31\pm0.08$ is implied, in excellent agreement with the figure
deduced from the cluster--galaxy cross--correlation function by Mo
\etal\ (1993).

We have calculated the distribution of the estimated long--wavelength
power, and searched for signs of non--Gaussianity in
the 1--point (in $k$--space) distribution of the power. We found no
trace of non--Gaussian behavior; rather, the distribution agreed
exceptionally well with the exponential distribution expected for
Gaussian fluctuations and we found no sign of periodicity or any
particularly strong spatial frequencies.
What is needed now is a well motivated non--Gaussian model with which to
compare this strong observational constraint.

\vskip 1cm
\noindent
{\bf Acknowledgements.} We would like to thank Karl Fisher, Michael
Vogeley and Anatoly Klypin for providing the IRAS 1.2-Jy, CfA and MDM
power spectra respectively. We would like to thank Andy Lawrence for
the updated QDOT data. HAF was supported in part by National
Science Foundation grant NFS--PHY--92--96020.

\def\ref{\par\noindent\hangindent=2pc \hangafter=1 }

\vskip 0.5cm
\noindent{\bf References}
\vskip 0.5cm

\ref Bardeen, J. M., Bond, J. R., Kaiser, N. \& Szalay, A. S. 1986, {\it ApJ}
{\bf 304} 15--61

\ref Baumgart, D. J. \& Fry, J. N. 1991, {\it ApJ} {\bf 375} 25--34

\ref Bond, J. R., \& Efstathiou G. 1984, {\it ApJ} {\bf 285} L45

\ref Broadhurst, T. J., Ellis, R. S., Koo, D. C. \& Szalay, A. S. 1990, {\it
Nature} {\bf 343} 726

\ref Cen, R., Gnedin, N. Y. Kofman, L. A. \& Ostriker, J. P. 1992, {\it ApJ
Lett} {\bf 399} L11

\ref Chester, T., Beichmann, C., \& Conrow, T. 1987, {\it Revised IRAS
Explanatory Supplement, Ch. XII}

\ref Efstathiou, G., Kaiser, N., Saunders, W., Lawrence, A.,
Rowan--Robinson, M., Ellis, R. S., \& Frenk, C. S. 1990, {\it MNRAS} {\bf 247}
10p

\ref Efstathiou, G., Bond, J. R. \& White, S. D.M.  1990, {\it MNRAS} {\bf 258}
1p

\ref Einasto, J., Gramann, M. \& Tago, E. 1993, {\it MNRAS} in press

\ref Feldman, H. A. 1993, {\it Texas/PASCOS Symposium} NY Academy of
Sciences Proceedings

\ref Fisher, K. B., Davis, M., Strauss, M. A., Yahil, A. \& Huchra,
J. P. 1993, {\it ApJ} {\bf 402} 42--57

\ref Hamilton, A. J. S., Kumar, P., Lu, E. \& Matthews, A., {\it ApJ} {\bf 374}
PL1

\ref Kaiser, N. 1987, {\it MNRAS} {\bf 227} 1

\ref Kaiser, N. 1992 ``The Density and Clustering of Mass in the
Universe''. Texas-ESO/CERN symposium

\ref Kaiser, N., Efstathiou, G., Ellis, R. S., Frenk, C. S., Lawrence, A.,
Rowan--Robinson, M. \& Saunders, W. 1991, {\it MNRAS} {\bf 252} 1

\ref Kaiser, N. \& Peacock, J. 1991, {\it ApJ} {\bf 379} 482

\ref Kendall, M. G. \& Stuart, A. 1977, {\it The advanced theory of statistics}
4th edition, (MacMillan Publishers, New York)

\ref Klypin, A., Holtzman, J., Primack, J. \& Reg\H os, E. 1992
Santa Cruz preprint SCIPP 92/52

\ref Mo, H. J., Peacock, J. A. \& Xia. X. Y. 1993, {\it MNRAS} {\bf 260}
121--131

\ref Moore, B., Frenk, C. S., Weinberg, D., Saunders, W.,
Lawrence, A., Ellis, R., Kaiser, N., Efstathiou, G., and
Rowan-Robinson, M. 1992., {\it MNRAS} {\bf 256}, 477

\ref Park, C., Gott, J. R. \& da Costa, L. N. 1992, {\it ApJ} {\bf 392}
L51--L54

\ref Peacock, J. A. 1991, {\it MNRAS} {\bf 253} 1p--5p

\ref Peacock, J. A. 1992, {\it New Insights into the Universe} Proceeding of
the
Valencia Summer school, ed. Martinez, V.J. (Dordrecht: Springer)

\ref Peacock, J. A. \& Nicholoson, D. 1991, {\it MNRAS} {\bf 253} 307--319

\ref Peacock, J. A. \& West, M.J. 1992, {\it MNRAS} {\bf 259} 494--504

\ref Peebles, P. J. E. 1973, {\it ApJ} {\bf 185} 413

\ref Peebles, P. J. E. 1980, {\it The Large Scale Structure of the
Universe}, Princeton University Press

\ref Peebles, P. J. E. \& Hauser, M. G. 1974, {\it ApJ Suppl.} {\bf 28} 19

\ref Rowan--Robinson, M. \etal\ 1990, {\it MNRAS} {\bf 247} 1

\ref Saunders, W. \etal\ 1991, {\it Nature} {\bf 349}, 32

\ref Shectman, S. A., Schechter, P. L, Oemler, A. A., Tucker, D., Kirshner, R.
P. \& Lin, H.
1992, {\it Clusters \& Superclusters of Galaxies}, ed. Fabian, A. C.,
(Kluwer), NATO ASI {\bf C366} 351

\ref Strauss, M. A., Davis, M. Yahil, A. \& Huchra, J. P. 1990, {\it ApJ} {\bf
361} 49

\ref Strauss, M. A., Yahil, A., Davis, M., Huchra, J. P. \& Fisher, K. 1992
IASSNS-AST 92/14

\ref Szalay, A. S., Ellis, R. S., Koo, D. C. \& Broadhurst, T. J. 1990, {\it
After the First
Three Minutes, AIP Conference Proceedings 222}, ed. Holt, Bennett \& Trimble,
(AIP,
New York) 261

\ref Vogeley, M. S., Park, C., Geller, M. \& Huchra, J. P. 1992, {\it Apj} {\bf
391} L5--L8

\ref Webster 1977,  {\it Radio Astronomy and Cosmology, IAU Symposium}
No {\bf 74} ed. Jauncey, D. L. (Reidel, Dordrecht)  75

\ref White, S. D. M., Efstathiou, G. \& Frenk, C. S. 1993, {\it MNRAS} in press

\ref Yu, J. T. \& Peebles, P. J. E. 1969, {\it ApJ} {\bf 158} 103

\vskip 0.5cm
\centerline{\bf Figure Captions}
\vskip 0.5cm

\item{Fig. 1)} The IRAS QDOT galaxy distribution (open circles) in the
range of $20\hm1 {\rm Mpc} < R < 500\hm1 {\rm Mpc}$ and the angular
mask (solid squares). Each masked bin is $\approx 1^o\times1^o$.

\item{Fig. 2)} A histogram of the IRAS QDOT survey and a $\chi^2$ fit of
Eqn. 2.5.2. (solid line) used as the radial selection function that
defines the synthetic catalogue and the mean space density of the
galaxies.

\item{Fig. 3)} The optimal weight function parameterized by the power
$P(k)$. By varying the assumed power over values that cover the
observed power, we in effect produce different effective survey depths.

\item{Fig. 4)} The power spectrum for the full IRAS QDOT survey
using four weight function parameterizations. We see
that the larger the power in the weight function is (\ie the greater
the effective depth of the survey) the more power we get.
The effect is most marked at the shallow end: we gain roughly
a factor 1.5 in power when the assumed power changes from
$2000\,(h^{-1}{\rm Mpc})^3$ to $4000\,(h^{-1}{\rm Mpc})^3$,
but things change relatively little thereafter. This suggests
that the effect is a local one that represents true sampling
fluctuations, but that the correct average power is detected
for greater depths.

\item{Fig. 5)} A comparison of the IRAS QDOT survey with $P(k)=2000$
($\hm1$ Mpc)$^3$ in the weight function with the 1.2--Jy IRAS survey.
Comparing with figure 3, we see that our shallowest weight gives
roughly constant weighting to volumes corresponding to the 1.2--Jy
survey, which is why we have presented the comparison in this way.
Once scaled to the same depth, the surveys agree reasonably well. The
1.2--Jy data lie slightly below our shallow results; however, there
would be a much more marked discrepancy if we had chosen to compare to
the `standard' $P=8000 (\hm1{\rm Mpc})^3$ QDOT results -- about a
factor of 1.5 in power.

\item{Fig. 6)} A comparison the IRAS QDOT survey with $P(k)=16000$
($\hm1$ Mpc)$^3$ in the weight function with the CfA surveys.  As with
QDOT, the CfA results appear to show some weak trend for the power to
increase with increasing sample depth.  The shapes of the CfA and QDOT
power spectra appear to differ: similar levels of power are seen at
large wavelength, but the CfA spectra show much more small--scale
power.  However, note the relative sizes of the large--wavelength
error bars in the CfA data, plus the fact that their analysis gives no
idea of the extent of any cross--correlation between different points.

\item{Fig. 7)} A comparison of the IRAS QDOT survey with $P(k)=8000$
($\hm1$ Mpc)$^3$ in the weight function with some theoretical models
normalized at $k=0.1h$ Mpc$^{-1}$.
1) For the BBKS linear CDM model with $\Omega h=0.5$, the poor fit is
apparent.  Whether one regards this as suggesting an excess of
large--scale power or a lack of small--scale power is a matter both of
taste and of where we choose to normalize.
2) The APM fitting function gives a reasonable fit to the IRAS QDOT
power spectrum. The overall shape agrees quite well with the APM data;
however, the position of the break that is required to best--fit the
APM data appears to be excluded by our results.
3) MDM simulations agree quite well with the IRAS QDOT data. The MDM
scenario gives results that are rather similar to those of CDM models
with low $\Omega h$ or `tilted' spectra.
4) $P(k)\propto k^{-1.4}$ gives good agreement for $k>0.04h$ Mpc$^{-1}$.

\item{Fig. 8)} Likelihood contours at $-\ln L = {\rm minimum} +
0.5,1,2\dots$.  The 95\% confidence
level would be at $\Delta\ln L=3$. This plot shows that the
maximum--likelihood model is well defined (and is an excellent
description of the data: $\chi^2=12$ on 18 degrees of freedom).

\item{Fig. 9)} The distribution of the power for $k<0.1$
($\lambda > 60\hm1$ Mpc). The agreement with the exponential
prediction is remarkable. It should be kept in mind that what we are
seeing here is the distribution of the `raw' power which contains a
superposition of the shot noise power, whose long wavelength Fourier
components is essentially guaranteed to be Gaussian distributed by
virtue of the central limit theorem. For the wavelengths employed
here, the real long--wavelength power is roughly equal to the shot
noise power.  The fact that the exponential law is nevertheless exact
out to $\hat P/\bar P\simeq 10$ is thus an extremely exacting test of
the Gaussian hypothesis -- and should provide an important constraint
on non--Gaussian models such as those based on topological defects.

\beginsection APPENDIX A

\beginsection Two-Point Function For a Poisson-Sample Point Process

If we have a point process $n(\br)$ which is a ``Poisson sample'' of
some continuous stochastic field $1 + f(\br)$ with a given mean
density of points $\nbar(\br)$ (i.e. the probability that an
infinitesimal volume element $\delta V$ contains an object is
$\nbar(\br)[1 + f(\br)] \delta V$) then the two-point function of
$n(\br)$ is
$$
\langle n(\br) n(\br') \rangle =
\nbar(\br) \nbar(\br') [1 + \xi(\br - \br')] + \nbar(\br) \delta(\br - \br')
\eqno(A1)
$$
where $\xi(\br) \equiv \langle f(\br') f(\br' + \br) \rangle$.

To prove this, consider the expectation value of
$$
\int d^3r\, \int d^3r'\, g(\br, \br') n(\br) n(\br')
\eqno(A2)
$$
where $g(\br, \br')$ is an arbitrary function. Using the standard
procedure (Peebles 1980, \S 36) of converting such integrals to sums
over infinitesimal microcells with occupation numbers $n_i = 0,1$ and
using $\langle n_i n_j \rangle = \nbar(\br_i) \nbar(\br_j) \delta V^2
[1 + \xi(\br_i - \br_j)]$ if $i \ne j$ and $\langle n_i^2 \rangle =
\langle n_i \rangle = \nbar(\br_i) \delta V$ we find
$$
\eqalign{
\left \langle \int d^3r\, \int d^3r'\,
	g(\br, \br') n(\br) n(\br') \right \rangle =&
	\int d^3r\, \int d^3r'\,
	g(\br, \br') \langle n(\br) n(\br') \rangle\cr
=& \sum_i\sum_j g(\br_i, \br_j) \langle n_i n_j \rangle\cr
}
$$
$$
\eqalign{
=& \sum_i\sum_j g(\br_i, \br_j) \nbar(\br_i) \nbar(\br_j)
[1 + \xi(\br_i - \br_j)] \delta V^2 +
\sum_i g(\br_i, \br_i) \nbar(\br_i) \delta V\cr
=& \int d^3r\, \int d^3r'\, g(\br, \br') \nbar(\br) \nbar(\br')
[1 + \xi(\br - \br')] + \int d^3r\, g(\br, \br) \nbar(\br)\cr
=& \int d^3r\, \int d^3r'\, g(\br, \br') \left\{
\nbar(\br) \nbar(\br') [1 + \xi(\br - \br')] + \nbar(\br) \delta(\br - \br')
\right\}\cr
}
\eqno(A3)
$$
Since this must be true for an arbitrary function $g(\br, \br')$ then
comparing the first and last lines of (A3) we obtain the identity (A1).

\beginsection APPENDIX B

\beginsection Fluctuations in power for a Gaussian field

We need to evaluate the two point function of fluctuations in the power
$\langle \delta\hat P({\bf k}) \delta\hat P({\bf k'}) \rangle$
[where $ \delta\hat P({\bf k})\equiv  \hat P(\bk) - P(k)$]:
$$
\langle \delta\hat P({\bf k}) \delta\hat P({\bf k'}) \rangle
= \langle \hat P({\bf k}) \hat P({\bf k'}) \rangle -
\langle P({\bf k}) \rangle \langle P({\bf k'}) \rangle.
\eqno(B1)
$$
To make progress, we shall assume that the large-wavelength
portion of the power spectrum describes a Gaussian field.
A possible way of proceeding would then be to consider
separately the real and imaginary parts of the Fourier
field $F({\bf k})=c_{\bf k} + i\,s_{\bf k}$, write down
all relevant correlations ($\langle c_{\bf k} s_{\bf k'}\rangle$
etc.) and use the general relation for a bivariate Gaussian
$$
\langle x^2 y^2 \rangle =\langle x^2 \rangle \langle y^2 \rangle
+3 \langle xy \rangle\,\sqrt{\langle x^2 \rangle \langle x^2 \rangle}
- \langle xy \rangle^2.
\eqno(B2)
$$
A less cumbersome method is to appeal to the idea that
realizations of Gaussian processes in $k$ space can be
obtained by Fourier transforming a set of independent
Gaussian random variables in real space:
$$
F({\bf k})=\sum_i g_i\, e^{i{\bf k\cdot r_i}}.
\eqno(B3)
$$
In these terms, the power $\langle |F({\bf k})|^2 \rangle=
\sum_i g_i^2$ and the two-point function in $k$ space is
$\langle F(\bk) F^*(\bk') \rangle=\sum_i g_i^2
\, e^{i{\bf (k-k')\cdot r_i}}$.
If we now write down the two-point function of the power
in terms of this expansion, we obtain a fourfold product
$\langle g_i g_j g_k g_\ell \rangle$, which only gives
a nonzero result when two pairs of indices are equal --
something that can happen in four distinct ways.
The two-point function for the power is then
$$
\langle \hat P({\bf k}) \hat P({\bf k'}) \rangle
= \sum_i \langle g_i^4 \rangle +
\sum_i\sum_{\ j\ne i}  \langle g_i^2 \rangle \, \langle g_j^2 \rangle \,
\left[ 1+
e^{i\bf(k+k')\cdot(r_i-r_j)} + e^{i\bf(k-k')\cdot(r_i-r_j)}
\right].
\eqno(B4)
$$
Since in the Gaussian case $\langle g_i^4 \rangle =
3 \langle g_i^2 \rangle^2$, this neatly allows the
double sum to be made unrestricted. If the term involving
$\bf k+k'$ is ignored on the grounds that its rapid
oscillations will give a result negligible by comparison
with the one involving $\bf k-k'$, then we obtain the desired
result:
$$
\langle \hat P({\bf k}) \hat P({\bf k'}) \rangle =
\langle P({\bf k}) \rangle \langle P({\bf k'}) \rangle +
|\langle F({\bf k}) F^*({\bf k'}) \rangle |^2.
\eqno(B5)
$$

\end